\documentclass[computation,review,submit,moreauthors,pdftex,10pt,a4paper]{mdpi} 

\firstpage{1} 
\articlenumber{x}
\doinum{10.3390/------}
\pubvolume{xx}
\pubyear{2018}
\copyrightyear{2018}
\externaleditor{Academic Editor: name}
\history{Received: date; Accepted: date; Published: date}
 \theoremstyle{mdpi}
 \newcounter{thm}
 \newcounter{ex}
 \newcounter{re}
 \theoremstyle{mdpidefinition}

\usepackage{subfig} 

\Title{Using the Maximum Entropy Principle to Combine Simulations and Solution Experiments}

\Author{Andrea Cesari $^{1}$, Sabine Rei{\ss}er $^{1}$ and Giovanni Bussi $^{1,}$*}
\AuthorNames{Andrea Cesari, Sabine Rei{\ss}er and Giovanni Bussi}

\address{%
$^{1}$ \quad Scuola Internazionale Superiore di
Studi Avanzati (SISSA), via Bonomea 265, 34136 Trieste, Italy}
\corres{Correspondence: bussi@sissa.it; Tel.: +39-040-3787407}

\abstract{
Molecular dynamics (MD) simulations allow investigating the structural dynamics of biomolecular systems with unrivaled time and space resolution.  However, in order to compensate for the inaccuracies of the utilized empirical  force fields, it is becoming common to integrate MD simulations with experimental data obtained from ensemble measurements.
 We here review the approaches that can be used to combine MD and experiment  under the guidance of the maximum entropy principle.
We mostly focus on methods based on Lagrangian multipliers, either implemented as reweighting of existing simulations or through an on-the-fly optimization. We discuss how errors in the experimental data can be modeled and accounted for.
Finally, we use simple model systems to illustrate the typical difficulties arising when applying these methods.
}

\preto{\abstractkeywords}{\nolinenumbers}
\begin{document}

\section{Introduction}
Molecular dynamics (MD) simulations are nowadays a fundamental tool used to complement experimental investigations in biomolecular modeling
\cite{dror2012biomolecular}. Although %
the accessible processes are usually limited to the microsecond timescale for classical MD with empirical force fields, with the help of enhanced sampling methods \cite{bernardi2015enhanced,valsson2016enhancing,mlynsky2017exploring} it
is possible to effectively sample events that would require a much longer time in order to spontaneously happen. However, the quality of the results is still limited by the accuracy of the employed force fields making experimental validations a necessary step.
The usual procedure consists in performing a simulation and computing some observable for which an experimental value has been already measured. If the calculated and experimental values are compatible, the simulation can be trusted and other observables can be estimated in order to make genuine predictions. If the discrepancy between calculated and experimental values is significant, one is forced to make a step back and perform a new simulation with a refined force field.
For instance, current force fields still exhibit
visible limitations in the study of protein-protein interactions~\cite{petrov2014current}, in the structural characterization of protein unfolded states~\cite{piana2011robust},
in the simulation of the conformational dynamics of unstructured RNAs~\cite{CondonTurner2015,BergonzoCheatham2015,sponer2018capturing},
and in the blind prediction of RNA structural motifs~\cite{Kuehrova2016,bottaro2016free,sponer2018capturing}.
However, improving force fields is a far-from-trivial task
because many correlated parameters should be adjusted. Furthermore, the employed functional forms might have an intrinsically limited capability to describe the real energy function of the system. Largely due to these reasons, it is becoming more and more common to restrain the simulations in order to enforce agreement with experimental data. Whereas this approach might appear not satisfactory, one should keep in mind that often experimental knowledge is anyway implicitly encoded in the simulation of complex systems (e.g., if the initial structure of a short simulation is taken from experiment, then the simulation will be biased toward it).
In addition, one should consider that validation can still be made against independent experiments
or against some of the data suitably removed from the set of restraints.
From another point of view,
the pragmatic approach of combining experiments with imperfect potential energy models
allows one to extract the maximum amount of information from sparse experimental data.
Particular care should be taken when interpreting bulk experiments that measure averages over a
large number of copies of the same molecule. These experiments are valuable in the characterization of dynamical molecules, where heterogeneous structures might be mixed and contribute with different weights to the experimental observation. If properly combined with MD
simulations, these experiments can be used to construct a high resolution picture of molecular structure and dynamics \cite{schroder2015hybrid,ravera2016critical,allison2017using,bonomi2017principles}.

In this review we discuss some recent methodological developments related to the application of the maximum entropy principle
to combine MD simulations with ensemble averages obtained from experiments (see, e.g., Refs.~\cite{PiteraChodera2012,Boomsma2014} for an introduction on this topic).
We briefly review the maximum entropy principle
and show how it can be cast into a minimization problem. We then discuss the equivalent formulation based on averaging between
multiple simultaneous simulations. Special explanations are dedicated to the incorporation of experimental errors in the
maximum entropy principle and to the protocols that can be used to enforce the experimental constraints.
Simple model systems 
are used to illustrate the typical
difficulties encountered in real applications. 
Source code for the model systems is available at \url{https://github.com/bussilab/review-maxent}.

\section{The Maximum Entropy Principle}

The maximum entropy principle dates back to 1957 when Jaynes \cite{jaynes1957information,jaynes1957information2} proposed it as a link between thermodynamic entropy
and information-theory entropy.
Previously, the definition of entropy was considered as an arrival point in the construction of new theories,
and only used as a validation against laws of thermodynamics \cite{jaynes1957information}.
In Jaynes formulation, maximum entropy was for the first time seen as the starting point to be used in building new theories.
In particular, distributions that maximize the entropy subject to some physical constraints 
were postulated to be useful in order to make inference on the system under study.
In its original formulation, the maximum entropy principle states that, given a system described by a number of states, the best probability distribution
for these states compatible with a set of observed data is the one maximizing the associated Shannon's entropy. This principle has been later extended to a maximum relative entropy principle
\cite{caticha2004relative} which has the advantage of being invariant with
respect to changes of coordinates and coarse-graining \cite{banavar2007maximum}
and has been shown to play an important role in multiscale problems \cite{shell2008relative}.
The entropy is here computed relative to a given prior distribution $P_0(\boldsymbol{q})$ and, for a system described by a set of continuous variables $\boldsymbol{q}$, is defined as

\begin{equation}
S[P||P_0]=-\int d\boldsymbol{q}\ P(\boldsymbol{q})\ln \frac{P(\boldsymbol{q})}{P_0(\boldsymbol{q})}~.
\end{equation}
This quantity should be maximized subject to constraints in order to be compatible with observations:

\begin{equation}
\label{eq:maxent}
\begin{cases}
P_{ME}(\boldsymbol{q}) = \underset{P(\boldsymbol{q})}{\arg \max}\ S[P||P_0]
\\
\int d\boldsymbol{q}\ s_{i}(\boldsymbol{q})P(\boldsymbol{q})=\langle s_{i}\left(\boldsymbol{q}\right)\rangle=s_{i}^{exp}\text{;} & i=1,\dots,M\\
\int d\boldsymbol{q}\ P(\boldsymbol{q})=1
\end{cases}
\end{equation}
Here $M$ experimental observations constrain the ensemble average of $M$ observables $s_i(\boldsymbol{q})$ computed over the distribution $P(\boldsymbol{q})$
to be equal to $s_i^{exp}$, and an additional constraint ensures that the distribution $P(\boldsymbol{q})$ is normalized.
$P_0(\boldsymbol{q})$ encodes the knowledge available before the experimental measurement and is thus called $\emph{prior}$ probability
distribution. $P_{ME}(\boldsymbol{q})$ instead represents the best estimate for the probability distribution after the experimental constraints have been enforced and is thus called \emph{posterior} probability distribution. Here the subscript $ME$ denotes the fact that this
is the distribution that maximizes the entropy.

Since the relative entropy $S[P||P_0]$ is the negative of the Kullback-Leibler divergence $D_{KL}[P||P_0]$
\cite{Kullback1951}, the procedure described above can be interpreted as a search for the posterior distribution that is as close as possible to the prior knowledge and agrees with the given experimental observations. In terms of information theory, the Kullback-Leibler divergence measures how much information is gained when prior knowledge $P_0(\boldsymbol{q})$ is replaced with $P(\boldsymbol{q})$.

The solution of the maximization problem in Eq.~\ref{eq:maxent} can be obtained using the method of Lagrangian multipliers, namely searching for the stationary points of the Lagrange function

\begin{equation}\mathcal{L}=S[P||P_0]-\sum_{i=1}^M
\lambda_i\left(\int d\boldsymbol{q}\ s_{i}(\boldsymbol{q})P(\boldsymbol{q})-s_i^{exp}\right)
-\mu 
\left(\int d\boldsymbol{q}\ P(\boldsymbol{q})-1\right)~,
\end{equation}
where $\lambda_i$ and $\mu$ are suitable Lagrangian multipliers.
The functional derivative of $\mathcal{L}$ with respect to $P(\boldsymbol{q})$ is

\begin{equation}
\frac{\delta \mathcal{L}}{\delta P(\boldsymbol{q})}=-\ln \frac{P(\boldsymbol{q})}{P_0(\boldsymbol{q})}-1-\sum_{i=1}^M\lambda_is_i(\boldsymbol{q})-\mu~.
\end{equation}
By setting $\frac{\delta \mathcal{L}}{\delta P(\boldsymbol{q})}=0$ and neglecting the normalization factor, the posterior reads

\begin{equation}
\label{eq:posterior}
P_{ME}(\boldsymbol{q})\propto e^{-\sum_{i=1}^M\lambda_is_i(\boldsymbol{q})}P_0(\boldsymbol{q})~.
\end{equation}
Here the value of the Lagrangian multipliers $\lambda_i$ should
be found by enforcing the agreement with the experimental data.
In the following, in order to have a more compact notation, we will drop the subscript from the Lagrangian multipliers and
write them as a vector whenever possible. Eq.~\ref{eq:posterior} could thus be equivalently written as

\begin{equation}
\label{eq:posterior2}
P_{ME}(\boldsymbol{q})\propto e^{-\boldsymbol{\lambda}\cdot\boldsymbol{s}(\boldsymbol{q})}P_0(\boldsymbol{q})~.
\end{equation}
Notice that the vectors $\boldsymbol{s}$ and $\boldsymbol{\lambda}$ have dimensionality $M$, whereas the vector $\boldsymbol{q}$ has dimensionality equal to the number of degrees of freedom of the analyzed system.

In short, the maximum relative entropy principle gives a recipe to obtain the
posterior distribution that is as close as possible to the prior distribution and agrees with some experimental observation.
In the following, we will drop the word ``relative'' and we will refer to this principle as the maximum entropy principle.

\subsection{Combining Maximum Entropy Principle and Molecular Dynamics\label{maxent-theory}}

When combining the maximum entropy principle with MD simulations the prior knowledge is represented by the probability distribution resulting from the employed potential energy, that is typically an empirical force field in classical MD.
In particular, given a potential energy $V_0(\boldsymbol{q})$, the associated probability distribution $P_0(\boldsymbol{q})$ at thermal equilibrium  is the Boltzmann distribution $P_0(\boldsymbol{q}) \propto e^{-\beta V_0(\boldsymbol{q})}$, where $\beta=\frac{1}{k_BT}$, $T$ is the system temperature, and $k_B$ is the Boltzmann constant.
According to Eq.~\ref{eq:posterior}, the posterior
will be $P_{ME}(\boldsymbol{q})
\propto
e^{-\boldsymbol{\lambda}\cdot\boldsymbol{s}(\boldsymbol{q})}
e^{-\beta V_0(\boldsymbol{q})}~.
$
The posterior distribution can thus be generated by a
modified potential energy in the form

\begin{equation}
\label{eq-v-me}
V_{ME}\left(\boldsymbol{q}\right)=V_0\left(\boldsymbol{q}\right)+k_BT\boldsymbol{\lambda}\cdot\boldsymbol{s}\left(\boldsymbol{q}\right)~.
\end{equation}
In other words, the effect of the constraint on the ensemble average is that of adding
a term to the energy that is linear in the function $\boldsymbol{s}(\boldsymbol{q})$ with
prefactors chosen in order to enforce the correct averages.
Such a linear term should be compared with constrained MD simulations, where the value of some
function of the coordinates is fixed at every step (e.g., using
the SHAKE algorithm \cite{ryckaert1977numerical}), or harmonic restraints, where a quadratic function of the observable is added to the potential energy function. Notice that the words constraint and restraint are usually employed when a quantity is exactly or softly enforced, respectively. Strictly speaking, in the maximum entropy context, ensemble averages
$\langle\boldsymbol{s}(\boldsymbol{q})\rangle$
are constrained whereas the corresponding functions $\boldsymbol{s}(\boldsymbol{q})$ are (linearly) restrained.

\begin{figure}
\centering
\includegraphics[width=\linewidth]{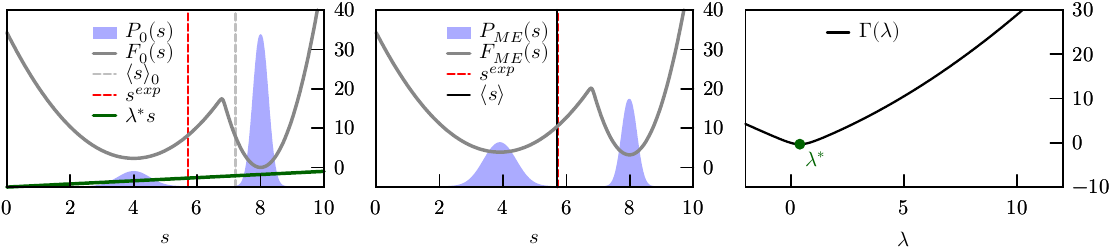}
\caption{The effect of a linear correcting potential on a given reference potential. 
$P_0(s)$ is the marginal probability distribution of some observable $s(\boldsymbol{q})$
according to the reference potential $V_0(\boldsymbol{q})$ and $F_0(s)$ is the corresponding free-energy profile (left panel).
Energy scale is reported in the vertical axis and is given in units of $k_BT$. Probability scales are not reported.
Vertical lines represent the average value of the observable $s$ in the prior ($\langle s \rangle_0$) and in the experiment ($s^{exp}$).
A correcting potential linear in $s$ (green line) shifts the relative depths of the two free-energy minima, leading to a new free energy
profile $F_{ME}(s)=F_0(s)+k_BT\lambda^*s$ that corresponds to a probability distribution $P_{ME}(s)$ (central panel). Choosing $\lambda^*$ equal to the value that minimizes $\Gamma(\lambda)$ (right panel) leads to an average $\langle s \rangle=s^{exp}$.
}
\label{linear-bias}
\end{figure}

If one considers the free energy as a function of the experimental
observables (also known as potential of mean force), which is defined as

\begin{equation}
F_0(\boldsymbol{s}')=-k_B T \ln \int d\boldsymbol{q}\  \delta(\boldsymbol{s}(\boldsymbol{q})-\boldsymbol{s}') P_0(\boldsymbol{q})~,
\end{equation}
the effect of the corrective potential in Eq.~\ref{eq-v-me} is just to tilt
the free-energy landscape

\begin{equation}
F_{ME}(\boldsymbol{s})=F_0\left(\boldsymbol{s}\right)+k_BT\boldsymbol{\lambda}\cdot\boldsymbol{s}+C~,
\end{equation}
where $C$ is an arbitrary constant.
A schematic representation of this tilting is reported in Fig. \ref{linear-bias}.

Any experimental data that is the result of an ensemble measurement can be used as a constraint.
Typical examples for biomolecular systems are nuclear-magnetic-resonance (NMR) experiments such as measures of chemical shifts \cite{case2013chemical}, scalar couplings \cite{Karplus1963}, or residual dipolar couplings \cite{tolman2006nmr}, and other techniques such as small-angle X-ray scattering (SAXS) \cite{bernado2007structural}, 
double electron-electron resonance (DEER)~\cite{jeschke2012deer},
and F\"orster resonance energy transfer~\cite{piston2007fluorescent}.
The only requirement is the availability of a so-called \emph{forward model} for such experiments. The forward model is a function mapping the atomic coordinates of the system to the measured quantity and thus allows the experimental data
to be back-calculated from the simulated structures.
For instance, in the case of 3J scalar couplings, the forward model
is given by the so-called Karplus relations~\cite{Karplus1963},
that are trigonometric functions of the dihedral angles.
It must be noted that the formulas used in standard forward models
are often parameterized empirically, and one should take into account  errors in these parameters on par with experimental errors (see Section~\ref{sec:modeling-errors}).
Without entering in the complexity of the methods mentioned above, we will only consider cases where experimental data can be trusted to be ensemble averages.

In short, the maximum entropy principle can be used to derive corrective potentials for MD simulations that constrain the value of some ensemble average. The choice to generate an ensemble that is as close as possible to the prior
knowledge implies that the correcting potential has a specific functional form, namely that it is linear in the
observables that have been measured.

\subsection{A Minimization Problem\label{minimization-sec}}

In order to chose the values of $\boldsymbol{\lambda}$ that satisfy Eq.~\ref{eq:maxent}, it is possible to recast the problem
into a minimization problem. In particular, consider the function~\cite{Mead1984,PiteraChodera2012}

\begin{equation}
\Gamma(\boldsymbol{\lambda})=\ln \left[ \int d\boldsymbol{q}\ P_{0}(\boldsymbol{q})e^{-\boldsymbol{\lambda}\cdot\boldsymbol{s}(\boldsymbol{q})}\right]
+\boldsymbol{\lambda}\cdot\boldsymbol{s}^{exp}~.
\label{gamma-func1-noerr}
\end{equation}
Notice that the first term is the logarithm of the ratio between the two partition functions associated to the potential energy functions
$V(\boldsymbol{q})$ and $V_0(\boldsymbol{q})$,
that is proportional to the free-energy difference between these two potentials.
The gradient of $\Gamma(\boldsymbol{\lambda})$ is

\begin{equation}
\frac{\partial \Gamma}{\partial \lambda_i}
=
s_i^{exp}-\frac{\int d\boldsymbol{q}\ P_{0}(\boldsymbol{q})e^{-\boldsymbol{\lambda}\cdot\boldsymbol{s}(\boldsymbol{q})}s_i(\boldsymbol{q})}{
\int d\boldsymbol{q}\ P_{0}(\boldsymbol{q})e^{-\boldsymbol{\lambda}\cdot\boldsymbol{s}(\boldsymbol{q})}}
=
s_i^{exp} -
\langle s_i(\boldsymbol{q})\rangle
\label{gamma-gradient}
\end{equation}
and is thus equal to zero when the average in the posterior distribution is identical to the enforced experimental value.
This means that the constraints in Eq.~\ref{eq:maxent} can be enforced by searching for a stationary point $\boldsymbol{\lambda}^*$ of $\Gamma(\boldsymbol{\lambda})$ (see Fig.~\ref{linear-bias}).
The Hessian of $\Gamma(\boldsymbol{\lambda})$ is 

\begin{equation}
\label{eq:hessian}
\frac{\partial \Gamma}{\partial \lambda_i \partial\lambda_j}
=
\langle s_i(\boldsymbol{q}) s_j(\boldsymbol{q}) \rangle
-
\langle s_i(\boldsymbol{q})\rangle \langle s_j(\boldsymbol{q}) \rangle
\end{equation}
and is thus equal to the covariance matrix of the forward models in the posterior distribution.
Unless the enforced observables are dependent on each other, the Hessian will be positive definite \cite{PiteraChodera2012}.
The solution of Eq.~\ref{eq:maxent} will thus correspond to a minimum of $\Gamma(\boldsymbol{\lambda})$
that can be searched for instance by a steepest descent procedure.
However there are cases where such minimum might not exist. In particular,
one should pay attention to the following cases:
\begin{itemize}
\item When data are incompatible with the prior distribution.
\item When data are mutually incompatible. As an extreme case, one can imagine two different experiments
that measure the same observable and report different values.
\end{itemize}
In both cases $\Gamma(\boldsymbol{\lambda})$ will have no stationary point. Clearly,
there is a continuum of possible intermediate situations where
data are almost incompatible.
In Section~\ref{sec:model-systems} we will see what happens when the
maximum entropy principle is applied to model systems designed in order to highlight these difficult situations.

\subsection{Connection with Maximum Likelihood Principle}

The function $\Gamma(\boldsymbol{\lambda})$ allows to easily highlight a connection between maximum entropy
and maximum likelihood principles.
Given an arbitrary set of $N_s$ molecular structures $\boldsymbol{q}_t$ chosen such
that $\frac{1}{N_s}\sum_{t=1}^{N_s} \boldsymbol{s}(\boldsymbol{q}_t) 
=\boldsymbol{s}^{exp}$,
it is possible to rewrite $e^{-N_s\Gamma(\boldsymbol{\lambda})}$ as 

\begin{equation}
\label{eq:exp-gamma}
e^{-N_s\Gamma(\boldsymbol{\lambda})}=
\frac{
e^{-N_s\boldsymbol{\lambda}\cdot
\boldsymbol{s}^{exp}
}
}{
\left[
\int d\boldsymbol{q} P_{0}(\boldsymbol{q})e^{-\boldsymbol{\lambda}\cdot\boldsymbol{s}(\boldsymbol{q})}
\right]^{N_s}
}
=
\frac{
e^{-\boldsymbol{\lambda}\cdot
\sum_t \boldsymbol{s}(\boldsymbol{q}_t)
}
}{
\left[
\int d\boldsymbol{q} P_{0}(\boldsymbol{q})e^{-\boldsymbol{\lambda}\cdot\boldsymbol{s}(\boldsymbol{q})}
\right]^{N_s}
}
=
\prod_{t=1}^{N_s}
\frac{
e^{-\boldsymbol{\lambda}\cdot\boldsymbol{s}(\boldsymbol{q}_t)}
}{
\int d\boldsymbol{q} P_{0}(\boldsymbol{q})e^{-\boldsymbol{\lambda}\cdot\boldsymbol{s}(\boldsymbol{q})}
}
=
\prod_{t=1}^{N_s}
\frac{
P(\boldsymbol{q}_t)
}{
P_{0}(\boldsymbol{q}_t)
}
\end{equation}
The last term is the ratio between the probability of drawing the structures $\boldsymbol{q}_t$ from the posterior distribution
and that of drawing the same structures from the prior distribution.
Since the minimum of $\Gamma(\boldsymbol{\lambda})$ corresponds to the maximum of $e^{-N_s\Gamma(\boldsymbol{\lambda})}$,
the distribution that maximizes the entropy under experimental constraints is identical to the one
that, among an exponential family of distributions, maximizes the likelihood of a set of structures
with average value of the observables $\boldsymbol{s}$ equal to the experimental value
\cite{berger1996maximum,Chen99agaussian}.
This equivalence can be considered as an added justification for the maximum entropy principle
\cite{berger1996maximum}: if the notion of selecting a posterior $P(\boldsymbol{q})$ that maximizes the entropy
is not compelling enough, one can consider that this same posterior is, among the distributions
with the exponential form of Eq.~\ref{eq:posterior}, the one that maximizes the likelihood of being compatible with
the experimental sample.

Equation~\ref{eq:exp-gamma} can also be rearranged to
$\Gamma(\boldsymbol{\lambda})=-\frac{1}{N_s}\sum_{t=1}^{N_s} \ln P(\boldsymbol{q}_t)
+\frac{1}{N_s}\sum_{t=1}^{N_s} \ln P_0(\boldsymbol{q}_t)$ and, after proper manipulation, it can be shown that
\begin{equation}
\label{eq:gamma-kl}
\Gamma(\boldsymbol{\lambda})=D_{KL}[P^{exp}||P] - D_{KL}[P^{exp}||P_0]
\end{equation}
where $P^{exp}$ is an arbitrary distribution with averages equal to the experimental ones.
Thus, minimizing $\Gamma(\boldsymbol{\lambda})$ is equivalent to choosing the distribution
with the exponential form of Eq.~\ref{eq:posterior} that is as close as possible to the experimental one.
Since at its minimum, by construction, $\Gamma(\boldsymbol{\lambda}^*)\le\Gamma(\boldsymbol{0})$,
it follows that
$
D_{KL}[P^{exp}||P_{ME}] \le D_{KL}[P^{exp}||P_0]
$. 
In other words, the maximum entropy restraint is guaranteed to make the posterior distribution closer to the experimental one than the prior distribution \cite{DannenhofferLafage2016}.

\subsection{Enforcing Distributions}
\label{sec:enforcing-distributions}

We so far considered the possibility of enforcing  ensemble averages.
However, one might be interested in enforcing the full distribution
of an observable. This can be  done by noticing that
the
marginal probability
distribution $\rho(\boldsymbol{s})$ of a quantity $\boldsymbol{s}$
can be computed as the expectation value of a Dirac-delta function:

\begin{equation}
\rho(\boldsymbol{s}')
=
\langle
\delta(\boldsymbol{s}(\boldsymbol{q})-\boldsymbol{s}')
\rangle~.
\end{equation}
An example of experimental technique that can report distance distributions
is the already mentioned DEER \cite{jeschke2012deer}.
If the form of $\rho(\boldsymbol{s})$ has been measured experimentally, the maximum entropy principle can be used 
to enforce it in a MD simulation. Notice that this corresponds to 
constraining an infinite number of data points (that is, the occupation
of each bin in the observable $\boldsymbol{s}$).
In this case, $\lambda$ will be a  function of $\boldsymbol{s}$
and 
Eq.~\ref{eq:posterior} will take the following form

\begin{equation}
\label{eq:posterior-ves}
P_{ME}(\boldsymbol{q})\propto 
e^{-\lambda(\boldsymbol{s}(\boldsymbol{q}))}
P_0(\boldsymbol{q})~.
\end{equation}
Thus, the correction to the potential should be a function of the observable $\boldsymbol{s}$
chosen in order to enforce the experimental distribution $\rho^{exp}(\boldsymbol{s})$.
Different approaches can be used to construct the function
$\lambda(\boldsymbol{s})$ with such property. For instance, one might take advantage of  iterative Boltzmann inversion procedures originally developed to derive coarse-grained models from atomistic simulations \cite{reith2003deriving}.
As an alternative, one might use a time-dependent adaptive potential.
In target metadynamics~\cite{white2015designing,marinelli2015ensemble} such potential is constructed as a sum of Gaussians
centered on the previously visited values of $\boldsymbol{s}$.
It can be shown that by properly choosing the prefactors of those
Gaussians an arbitrary target distribution can be enforced.

Alternatively, it is possible to directly minimize the function $\Gamma(\boldsymbol{\lambda})$ as mentioned in  Section \ref{minimization-sec}. In this context, $\Gamma$ would be a functional of $\lambda(\boldsymbol{s})$ with the form

\begin{equation}
\Gamma[\lambda]=\ln \int d\boldsymbol{q}
\ 
e^{-\lambda(\boldsymbol{s}(\boldsymbol{q}))}
P_0(\boldsymbol{q})
+\int d\boldsymbol{s}
\lambda(\boldsymbol{s})\rho^{exp}(\boldsymbol{s})~.
\end{equation}
Interestingly, this functional is identical to the one introduced
in the variationally enhanced sampling (VES) method of Ref.~\cite{Valsson2014}.
In its original formulation, VES was
used to enforce a flat distribution in order to sample rare events.
However, the method
can also be used to enforce an arbitrary a priori chosen
distribution~\cite{ShafferValssonParrinello2016,invernizzi2017coarse}. The analogy with maximum entropy methods,
together with the relationship in Eq.~\ref{eq:gamma-kl},  was already noticed in Ref.~\cite{invernizzi2017coarse} and is interesting for a twofold reason:
(a) It provides a maximum-entropy interpretation of VES and
(b) the numerical techniques used for VES might be used to enforce experimental averages in a maximum-entropy context.
We will further comment about this second point in Section~\ref{sec:other-on-the-fly}.

\subsection{Equivalence to the Replica Approach}
\label{sec:replica}
A well established method to enforce ensemble averages in molecular simulations is represented by restrained ensembles \cite{fennen1995structure,BestVendruscolo2004,LindorffLarsenBestVendruscolo2005}. The rationale behind this method is to mimic an ensemble of structures by simulating in parallel $N_{rep}$ multiple identical copies (replicas) of the system each of which having its own atomic coordinates. The agreement with the $M$ experimental data is then enforced by adding a harmonic restraint for each observable, centered on the experimental reference and acting on the average over all the simulated replicas. This results in a restraining potential with the following form:

\begin{equation}
\label{eqn:replicas}
V_{RE}\left(\boldsymbol{q}_1,\boldsymbol{q}_2,\dots,\boldsymbol{q}_{N_{rep}}\right)=\sum_{i=1}^{N_{rep}} V_{0}\left(\boldsymbol{q}_i\right)+\frac{k}{2}\sum_{j=1}^{M}\left(\frac{1}{N_{rep}}\sum_{i=1}^{N_{rep}}s_{j}(\boldsymbol{q}_i)-s_{j}^{exp}\right)^{2}~,
\end{equation}
where $k$ is a suitably chosen force constant. It has been shown \cite{PiteraChodera2012,CavalliCamilloniVendruscolo2013,RouxWeare2013} that this method produces the same ensemble as the maximum entropy approach in the limit of  large number of replicas
$\left(N_{rep}\rightarrow\infty\right)$.
Indeed, the potential in Eq.~\ref{eqn:replicas} results in  the same force
$-\frac{k}{N_{rep}}\left(\frac{1}{N_{rep}}\sum_{i=1}^{N_{rep}}s_{j}(\boldsymbol{q}_i)-s_{j}^{exp}
\right)$ applied to the observable $s_{j}(\boldsymbol{q})$ in each replica.
As the number of replicas grows, the fluctuations of the average decrease and the applied force becomes constant in time,
so that the explored distribution will have the same form as Eq.~\ref{eq:posterior} with
$\boldsymbol{\lambda}=\frac{k}{N_{rep}k_BT}\left(\frac{1}{N_{rep}}\sum_{i=1}^{N_{rep}}\boldsymbol{s}(\boldsymbol{q}_i)-\boldsymbol{s}^{exp} \right)$.
If $k$ is  chosen large enough, the average between the replicas will be forced to be equal to the experimental
one. It is possible to show that, in order to enforce the desired average, $k$ should grow faster than $N_{rep}$
\cite{RouxWeare2013}.
In practical implementations,
$k$ should be finite in order to avoid infinite forces.
A direct calculation of the entropy-loss due to the choice of a finite $N_{rep}$ has been proposed
to be an useful tool in the search for the correct number of replicas \cite{olsson2015quantification}.
An approach based on a posteriori reweighting (Section~\ref{sec:reweighting}) of
replica-based simulations and named Bayesian inference of ensembles has been also
proposed in order to eliminate the effect of choosing a finite
number of replicas \cite{Hummer2015}.

\section{Modelling Experimental Errors\label{modelling-exp-errors}}
\label{sec:modeling-errors}
The maximum entropy method can be modified in order to account for uncertainties in experimental data. This step is fundamental in order to reduce over-fitting. In this section we will briefly consider how the error can be modeled according to Ref.~\cite{cesari2016combining}. 
Here errors are modeled modifying the experimental constraints introduced in Eq.~\ref{eq:maxent} by introducing an auxiliary variables $\epsilon_i$ for each data point representing
the discrepancy or residual between the experimental and the simulated value. The new constraints are hence defined as follows:

\begin{equation}
\langle\left(\boldsymbol{s}(\boldsymbol{q})+\boldsymbol{\epsilon}\right)\rangle=\boldsymbol{s}^{exp}~.
\label{constrains-with-eps}
\end{equation}
The auxiliary variable $\boldsymbol{\epsilon}$ is a vector with dimensionality equal to the number of constraints and models all the possible sources of error, including inaccuracies of the forward models (Section \ref{maxent-theory}) as well as experimental uncertainties. Errors can be modeled by choosing a proper prior distribution function for the variable $\boldsymbol{\epsilon}$. A common choice is represented by a Gaussian prior with a fixed standard deviation $\sigma_i$ for the $i^{th}$ observable

\begin{equation}
P_0\left(\boldsymbol{\epsilon}\right)\propto\prod_{i=1}^M\exp\left(-\frac{\epsilon_{i}^{2}}{2\sigma_{i}^{2}}\right)~.
\label{gaussian-prior}
\end{equation}
The value of $\sigma_i$ corresponds to the level of confidence in the $i^{th}$ data point, where $\sigma_i=\infty$ implies to completely discard the data in the optimization process while $\sigma_i=0$ means having complete confidence in the data, that will be fitted as best as possible. Notice that, for additive errors, $\boldsymbol{q}$ and $\boldsymbol{\epsilon}$ are independent variables and Eq.~\ref{constrains-with-eps} can be written as:

\begin{equation}
\langle \boldsymbol{s}(\boldsymbol{q})\rangle=\boldsymbol{s}^{exp}-\langle\boldsymbol{\epsilon}\rangle
\label{new-target}
\end{equation}
where $\langle\boldsymbol{\epsilon}\rangle$ is computed in the posterior distribution $P(\boldsymbol{\epsilon})\propto P_0(\boldsymbol{\epsilon})e^{-\boldsymbol{\lambda}\cdot\boldsymbol{\epsilon}}$. Incorporating the experimental error in the maximum entropy approach is then as easy as enforcing a different experimental value, corresponding to the one in Eq.~\ref{new-target}. Notice that the value of $\langle\boldsymbol{\epsilon}\rangle$ only depends on its prior distribution $P_0(\boldsymbol{\epsilon})$ and on $\boldsymbol{\lambda}$.
For a Gaussian prior with standard deviation $\sigma_i$ (Eq.~\ref{gaussian-prior}) we have:

\begin{equation}
\langle\epsilon_i\rangle=-\lambda_i\sigma_i^{2}
\label{gauss-epsilon}.
\end{equation}
Thus, as $\lambda$ grows in magnitude, a larger discrepancy between simulation and experiment will be accepted.
In addition, it can be seen that applying the same constraint \emph{twice} is exactly equivalent to applying a constraint
with a $\sigma^2_i$ reduced by a factor two. This is consistent with the fact that the confidence in the repeated data point is increased.

Other priors are also possible in order to better account for outliers
and to deal with cases where the standard deviation of the residual
is not known a priori. One might consider the variance of the $i^{th}$ residual $\sigma^2_{0,i}$ as a variable sampled from a given prior distribution $P_0(\sigma^2_{0,i})$:

\begin{equation}
P_{0}\left(\boldsymbol{\epsilon}\right)=\prod_{i=1}^M\int_{0}^{\infty}d\sigma^2_{0,i}P_0(\sigma^2_{0,i})\frac{1}{\sqrt{2\pi\sigma^2_{0,i}}}\exp\left(-\frac{\epsilon_{i}^{2}}{2\sigma_{0,i}^2}\right)~.
\label{generic-prior}
\end{equation}
A flexible functional form for $P_0(\sigma^2_{0,i})$ can be obtained
using the following Gamma distribution 

\begin{equation}
P_{0}(\sigma_{0,i}^2)\propto(\sigma^2_{0,i})^{\kappa-1}\exp\left(-\frac{\kappa\sigma_{0,i}^{2}}{\sigma_i^{2}}\right).
\label{generic-prior2}
\end{equation}
In the above equation $\sigma^2_i$ is the mean parameter of the Gamma function and must be interpreted as the typical expected variance of the error on the $i^{th}$ data point.
$\kappa$, which must satisfy $\kappa>0$, is the shape parameter of the Gamma distribution and expresses how much the distribution is peaked around $\sigma_{i}^2$. In practice, it controls how much the optimization is tolerant to large discrepancies between the experimental data and the enforced average. Notice that in Ref.~\cite{cesari2016combining} a different convention was used with a parameter $\alpha=2\kappa-1$.
By setting $\kappa=\infty$ a Gaussian prior on $\boldsymbol{\epsilon}$ will be recovered. Smaller values of $\kappa$ will lead to a prior distribution on $\boldsymbol{\epsilon}$ with ``fatter'' tails and thus able to accommodate larger differences between experiment and simulation.
For instance, the case $\kappa=1$ leads to a Laplace prior
$P_0(\boldsymbol{\epsilon})\propto \prod_i \exp\left(-\frac{\sqrt{2}\left|\boldsymbol{\epsilon}\right|}{\sigma_i}\right)$.
After proper manipulation, the resulting expectation
value $\langle\boldsymbol{\epsilon}\rangle$ can be shown to be

\begin{equation}
\langle\epsilon_{i}\rangle=
-\frac{\lambda_i\sigma_i^2}{1-\frac{\lambda_i^2\sigma_i^2}{2\kappa}}~.
\end{equation}
In this case, it can be seen that applying the same constraint twice is exactly equivalent to applying a constraint
with a $\sigma^2_i$ reduced by a factor two and a $\kappa$ multiplied by a factor two.

In terms of the minimization problem of Section \ref{minimization-sec},
modeling experimental errors as discussed here is equivalent to adding a contribution $\Gamma_{err}$ to Eq.~\ref{gamma-func1-noerr}:

\begin{equation}
\Gamma(\boldsymbol{\lambda})=\ln \int d\boldsymbol{q}\ P_{0}(\boldsymbol{q})e^{-\boldsymbol{\lambda}\cdot\boldsymbol{s}(\boldsymbol{q})}
+\boldsymbol{\lambda}\cdot\boldsymbol{s}^{exp}
+ \Gamma_{err}(\boldsymbol{\lambda})~.
\end{equation}
For a Gaussian noise with preassigned variance (Eq.~\ref{gaussian-prior}) the additional term is

\begin{equation}
\Gamma_{err}(\boldsymbol{\lambda})=\frac{1}{2}\sum_{i=1}^M \lambda^2_i\sigma^2_i~.
\label{gamma-func-err}
\end{equation}
For a prior on the error in the form of Eqs.~\ref{generic-prior} and \ref{generic-prior2} one obtains

\begin{equation}
\Gamma_{err}(\boldsymbol{\lambda})= -\kappa \sum_{i=1}^M\ln\left(1-\frac{\lambda_i^{2}\sigma_i^{2}}{2\kappa}\right)~.
\label{gamma-func-err-generic}
\end{equation}
In the limit of large $\kappa$, Eq.~\ref{gamma-func-err-generic} is equivalent to Eq.~\ref{gamma-func-err}.
If the data points are expected to all have the same error $\sigma_0$,
unknown but with a typical value $\sigma$ (see Ref.~\cite{cesari2016combining}), Eq.~\ref{gamma-func-err-generic} should be modified to
$
\Gamma_{err}(\boldsymbol{\lambda})= -\kappa \ln\left(1-\frac{\left|\boldsymbol{\lambda}\right|^2\sigma^{2}}{2\kappa}\right)
$.

Equation~\ref{gamma-func-err-generic} shows that by construction the Lagrangian multiplier $\lambda_i$ will be limited in the range
$\left(-\frac{\sqrt{2\kappa}}{\sigma_i},\ +\frac{\sqrt{2\kappa}}{\sigma_i}\right)$.
The effect of using a prior with $\kappa<\infty$ is thus that of
restricting the range of allowed $\lambda$ in order to avoid too large
modifications of the prior distribution.
In practice, values of $\lambda$ chosen outside these boundaries would
lead to a posterior distribution
$P(\boldsymbol{\epsilon})\propto P_0(\boldsymbol{\epsilon})
e^{-\boldsymbol{\lambda}\cdot\boldsymbol{\epsilon}}$ that cannot be normalized.

Except for trivial cases (e.g., for Gaussian noise with $\sigma=0$), the contribution originating from error modeling has positive definite
Hessian and as such it makes $\Gamma(\lambda)$ a strongly convex function. Thus, a suitable error treatment can make the 
minimization process numerically easier.

It is worth mentioning that a very similar formalism can be used
to include not only errors but more generally any quantity that influences the experimental measurement but cannot be directly obtained from the simulated structures. For instance, in the case of residual dipolar couplings \cite{tolman2006nmr},
the orientation of the considered molecule with the respect to the external field is often unknown. The orientation of the
field can then be used as an additional vector variable to be sampled with
a Monte Carlo procedure, and suitable Lagrangian multipliers can
be obtained in order to enforce the agreement with experiments \cite{olsson2015molecular}. Notice that in this case the orientation
contributes to the ensemble average in a non-additive manner so that
Eq.~\ref{new-target} cannot be used.
Interestingly,
thanks to the equivalence between multi-replica simulations
and maximum entropy restraints (Section~\ref{sec:replica}), equivalent results can be obtained
using the tensor-free method of Ref.~\cite{camilloni2014tensor}.

Finally, we note that several works introduced error treatment using a Bayesian framework
\cite{Das2014,Hummer2015,BonomiMETAINF2016,BrookesHead-Gordon2016}. Interestingly, Bayesian ensemble refinement~\cite{Hummer2015} introduces an additional parameter ($\theta$) that takes into account the confidence in the prior distribution. In case of Gaussian error, this parameter enters as a global scaling factor in the errors $\sigma_i$ for each data point.
Thus, the errors $\sigma_i$ discussed above can be used to modulate both our confidence in experimental data and our confidence in the original force field.
The equivalence between the error treatment of Ref.~\cite{Hummer2015} and the one reported here is further discussed in Ref.~\cite{cesari2016combining},
in particular for what concerns non-Gaussian error priors.

\section{Exact Results on Model Systems}
\label{sec:model-systems}

In this section we illustrate the effects of adding restraints using the maximum entropy principle on simple
model systems.
In order to do so we first derive some simple relationship valid
when the prior has a particular functional form, namely a sum of $N_G$ Gaussians with center $\boldsymbol{s}_{\alpha}$
and covariance matrix $A_{\alpha}$, where $\alpha=1,\dots,N_G$:

\begin{equation}
P_0(\boldsymbol{s})=\sum_{\alpha=1}^{N_G} \frac{w_{\alpha}}{\sqrt{2\pi\det A_{\alpha}}} e^{-\frac{(\boldsymbol{s}-\boldsymbol{s}_{\alpha})
A^{-1}_{\alpha} (\boldsymbol{s}-\boldsymbol{s}_{\alpha})}{2}}~.
\end{equation}
The coefficients $w_{\alpha}$ provide the weights of each Gaussian and are normalized ($\sum_{\alpha}w_{\alpha}=1$).
We here assume that the restraints are applied on the variable $\boldsymbol{s}$. For a general system, one should first perform
a dimensional reduction in order to obtain the marginal prior probability $P_0(\boldsymbol{s})$.
By constraining the ensemble averages of the variable $\boldsymbol{s}$ to an experimental value $\boldsymbol{s}^{exp}$
the posterior becomes:

\begin{equation}
P_{ME}(\boldsymbol{s})= \frac{e^{-\boldsymbol{\lambda} \cdot \boldsymbol{s} } }{Z(\boldsymbol{\lambda})}
\sum_{\alpha} \frac{w_{\alpha}}{\sqrt{2\pi\det A_{\alpha}}} e^{-\frac{(\boldsymbol{s}-\boldsymbol{s}_{\alpha}) A^{-1}_{\alpha} (\boldsymbol{s}-\boldsymbol{s}_{\alpha})}{2}}~.
\end{equation}
With proper algebra it is possible to compute explicitly the normalization factor
$
Z(\boldsymbol{\lambda})=\sum_{\alpha}
w_{\alpha}e^{\frac{\boldsymbol{\lambda} A_{\alpha} \boldsymbol{\lambda} }{2} - \boldsymbol{\lambda} \cdot \boldsymbol{s}_{\alpha}}
$.
The function $\Gamma(\boldsymbol{\lambda})$ to be minimized is thus equal to:

\begin{equation}
\Gamma(\boldsymbol{\lambda})=
\ln\left(
\sum_{\alpha}
w_{\alpha}e^{\frac{\boldsymbol{\lambda} A_{\alpha} \boldsymbol{\lambda} }{2} - \boldsymbol{\lambda} \cdot \boldsymbol{s}_{\alpha}}
\right)
+\boldsymbol{\lambda} \cdot \boldsymbol{s}^{exp}
+\Gamma_{err}(\boldsymbol{\lambda})
\end{equation}
and the average value of $\boldsymbol{s}$ in the posterior is

\begin{equation}
\langle\boldsymbol{s}\rangle=
\frac{\sum_{\alpha}
w_{\alpha}e^{\frac{\boldsymbol{\lambda} A_{\alpha} \boldsymbol{\lambda} }{2} - \boldsymbol{\lambda} \cdot \boldsymbol{s}_{\alpha}}
\left(\boldsymbol{s}_{\alpha}
-A_{\alpha}\boldsymbol{\lambda} 
\right)}{\sum_{\alpha}
w_{\alpha}e^{\frac{\boldsymbol{\lambda} A_{\alpha} \boldsymbol{\lambda} }{2} - \boldsymbol{\lambda} \cdot \boldsymbol{s}_{\alpha}}}~.
\end{equation}
We could not find a 
close formula for $\boldsymbol{\lambda}^*$ given $\boldsymbol{s}^{exp}$ and $\Gamma_{err}$.
However, the solution can be found numerically with the gradient descent procedure discussed in
Section \ref{sec-strategies} (see Eq.~\ref{gradient-descent}).

\subsection{Consistency between Prior Distribution and Experimental Data}
\begin{figure}
\centering
\includegraphics[width=\linewidth]{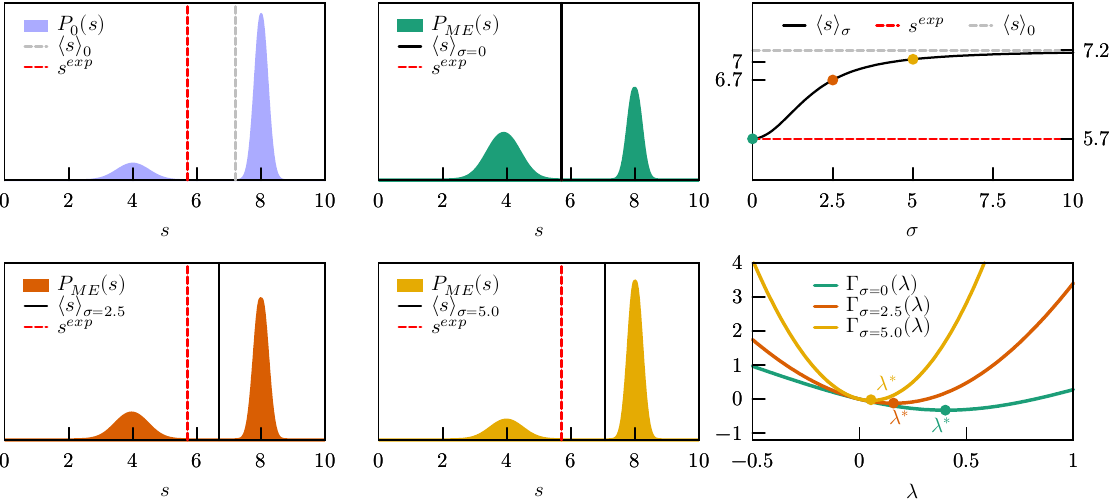}
\caption{Effect of modeling error with a Gaussian probability distribution with different standard deviations $\sigma$ on the posterior distribution $P_{ME}(s)$.
The experimental value is here set to $s^{exp}=5.7$, which is compatible with the prior distribution. Left and middle column: prior $P_0(s)$ and posterior $P_{ME}(s)$ with $\sigma=0,\ 2.5,\ 5.0$. Right column: ensemble average $\langle s\rangle$ plotted as a function of $\sigma$ and $\Gamma(\lambda)$ plotted for different values of $\sigma$.
$\lambda^*$ denotes that value of $\lambda$ that minimizes $\Gamma(\lambda)$.}
\label{sigma-gamma57}
\end{figure}
\begin{figure}
\centering
\includegraphics[width=\linewidth]{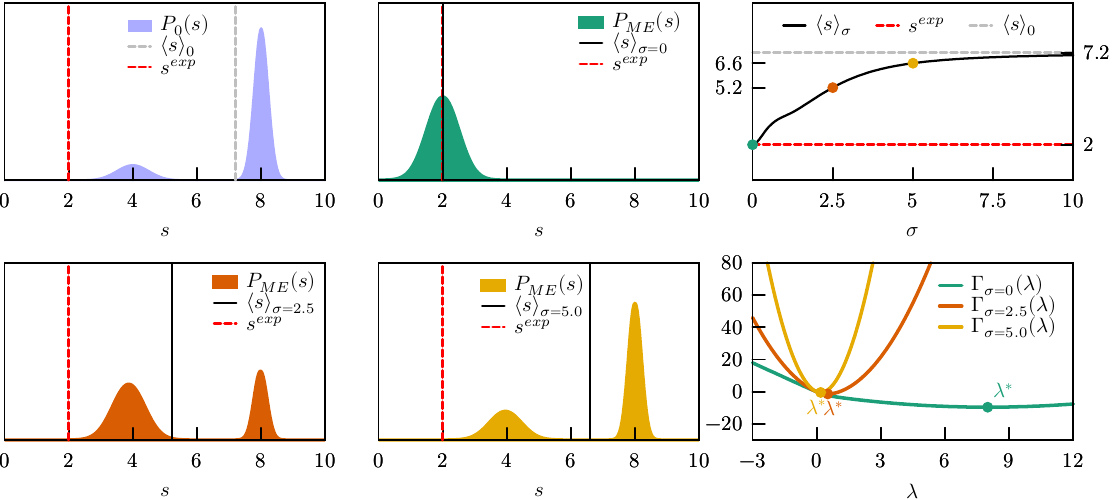}
\caption{Same as Fig.~\ref{sigma-gamma57}, but the experimental value  is here set to $s^{exp}=2$, which is almost incompatible with the prior distribution.}
\label{sigma-gamma}
\end{figure}
We consider a one dimensional model with a prior expressed as a sum of two Gaussians, one 
centered in $s_A=4$ with standard deviation $\sigma_A=0.5$ and one 
centered in $s_B=8$ with standard deviation $\sigma_B=0.2$.
The weights of the two Gaussians are $w_A=0.2$ and $w_B=0.8$, respectively.
The prior distribution is thus 
$P_0(s)\propto \frac{w_A}{\sigma_A}e^{-(s-s_A)^2/2\sigma_A^2} +
\frac{w_B}{\sigma_B} e^{-(s-s_B)^2/2\sigma_B^2}$, has an average value $\langle s\rangle_0=7.2$, and is represented in Fig. \ref{sigma-gamma57}, left column top panel.

We first enforce a value $s^{exp}=5.7$, which is compatible with the prior probability.
If we are absolutely sure about our experimental value and set $\sigma=0$, the $\lambda^*$ which minimizes $\Gamma(\lambda)$ is $\lambda^*\approx 0.4$ (Fig. \ref{sigma-gamma57} right column, bottom panel).
In case values of $\sigma \neq 0$ are used,
the $\Gamma(\boldsymbol{\lambda})$ function becomes more convex and the
optimal value $\lambda^*$ is decreased. As a result, the
average $s$ in the posterior distribution is approaching its value in the prior.
The evolution of the ensemble average $\langle s\rangle_{\sigma}$ with $\sigma$  values between zero and ten, with respect to the initial $\langle s\rangle_{0}$ and the experimental $s^{exp}$, is shown in Fig. \ref{sigma-gamma57}, right column top panel. 
In all these cases the posterior distributions remain bimodal
and the main effect of the restraint is to change the relative population
of the two peaks (Fig. \ref{sigma-gamma57}, left and middle columns).

We then enforce an average value $s^{exp}=2$, which is far outside the original probability distribution (see Figure \ref{sigma-gamma}). 
If we are absolutely sure about our experimental value and set $\sigma=0$, the $\lambda^*$ which minimizes $\Gamma(\lambda)$ is very large, $\lambda^*\approx 8$ (Fig. \ref{sigma-gamma} right column, bottom panel).
Assuming zero error on the experimental value is equivalent to having poor confidence in the probability distribution sampled by the force field, and leads in fact to a $P_{ME}(s)$ completely different from $P_{0}(s)$.
The two peaks in $P_{0}(s)$ are replaced by a single peak centered around the experimental value, which is exactly met by the ensemble average ($\langle s\rangle_{\sigma=0}=s^{exp}=2$; Fig. \ref{sigma-gamma} middle column top panel). 
Note that this is possible only because the experimental value is not entirely incompatible with the prior distribution, i.e. it has a small, non-zero probability also in the prior. If the probability had been zero, $\Gamma(\lambda)$ would have had no minimum and no optimal $\lambda^*$ would have been found.
If we have more confidence in the distribution sampled by the force field, assume that there might be an error in our experimental value, and set $\sigma=2.5$, $\lambda^*$ is more than one order of magnitude lower ($\lambda^*\approx 0.52$). 
The two peaks in $P_{0}(s)$ are only slightly shifted towards lower $s$, while their relative populations are shifted in favor of the peak  centered around 4 (Fig. \ref{sigma-gamma}, left column bottom panel).  
According to our estimate of the probability distribution of the error, the ensemble average $\langle s\rangle_{\sigma=2.5}\approx 5.2$ is more probably the true value than the experimentally measured one.
In case we have very high confidence in the force field and very low confidence in the experimental value and set $\sigma=5.0$, the correction becomes very small
($\lambda^*\approx 0.18$) and the new ensemble average $\langle s\rangle_{\sigma=5.0}\approx 6.6$, very close to the initial $\langle s\rangle_{0}=7.2$ (Fig. \ref{sigma-gamma}, middle column bottom panel).
The evolution of the ensemble average $\langle s\rangle_{\sigma}$ with $\sigma$  values between zero and ten, with respect to the initial $\langle s\rangle_{0}$ and the experimental $s^{exp}$, is shown in Fig. \ref{sigma-gamma}, right column top panel.

In conclusion, when data that are not consistent with the prior distribution are enforced, the posterior distribution could be severely distorted.
Clearly, this could happen either because the prior is completely wrong or
because the experimental values are affected by errors. By including a suitable error model in the maximum entropy procedure it is possible to easily interpolate between the two extremes in which we completely trust the force field or the experimental data.

\subsection{Consistency between Data Points}
\label{sec-2d-systems}

\begin{figure}
\centering
\includegraphics[width=\linewidth]{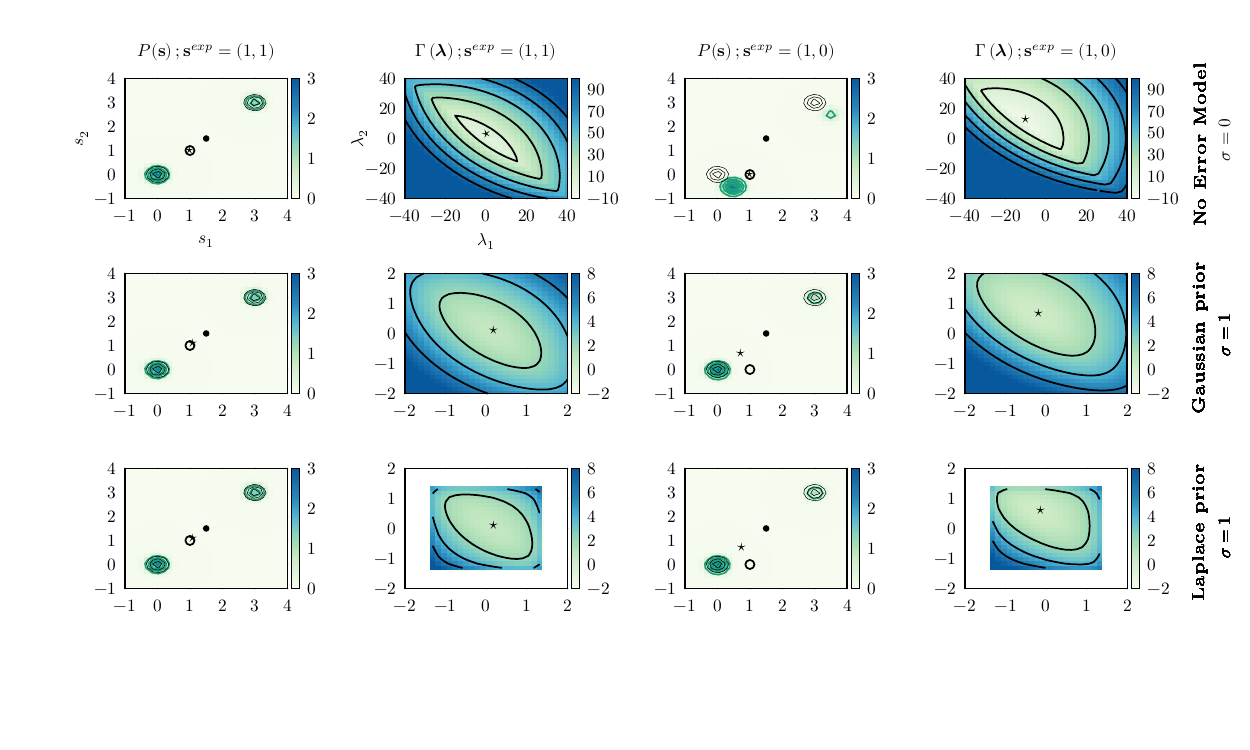}

\caption{Effect of different prior distributions for the error model in a two-dimensional system.
In the first (last) two columns, compatible (incompatible) data are enforced.
In the first and the third column, prior distributions are represented as black contour lines and posterior
distributions are shown in color scale. A black dot and a $\star$ are used to indicate the average values of $\boldsymbol{s}$ in the prior and posterior
distributions respectively, while an empty circle is used to indicate the target $\boldsymbol{s^{exp}}$. 
In the second and the fourth column, the function $\Gamma(\boldsymbol{\lambda})$
is shown, and its minimum $\boldsymbol{\lambda}^*$ is indicated with a $\star$. The first row reports results where errors are not modeled, whereas the second and the third row report
results obtained using Gaussian and Laplace prior for the error model respectively.
Notice that a different scale is used to represent $\Gamma(\boldsymbol{\lambda})$ in the first row.
For the Laplace prior, the region of $\boldsymbol{\lambda}$ where $\Gamma(\boldsymbol{\lambda})$ is undefined
is marked as white.
}
\label{2d-model}
\end{figure}

We then consider a two dimensional model with a prior expressed as a sum of two Gaussians centered in
$\boldsymbol{s}_A=(0,0)$ and $\boldsymbol{s}_B=(3,3)$ with identical standard deviations $\sigma_A=\sigma_B=0.2$ and weights $w_A=w_B=0.5$.
The prior distribution is
represented in Fig.~\ref{2d-model}.
This model is particularly instructive since, by construction, the two components of $\boldsymbol{s}$  are highly correlated and is hence possible to see what happens when inconsistent data are enforced. To this aim we study the two scenarios (i.e., consistent and inconsistent data) using different error models (no error model, Gaussian prior with $\sigma=1$, and Laplace prior with $\sigma=1$), for a total of six combinations.
In the \textit{consistent} case we enforce $\boldsymbol{s}^{exp}=(1,1)$,
whereas in the \textit{inconsistent} one we enforce
$\boldsymbol{s}^{exp}=(1,0)$.
Figure \ref{2d-model} reports the posterior distributions obtained
in all these cases.

When consistent data are enforced
the posterior distribution is very similar to the prior distribution,
the only difference being a modulation in the weights of the two peaks.
The optimal value $\boldsymbol{\lambda}^*$, marked with a $\star$ in Figure \ref{2d-model}, does not depend significantly
on the adopted error model.
The main difference between including or not including error models can be seen in the form of the $\Gamma(\boldsymbol{\lambda})$ function. When errors are not included, $\Gamma(\boldsymbol{\lambda})$ is almost flat in a given direction, indicating that one of the eigenvalues of its Hessian is very small. On the contrary,
when error modeling is included, the $\Gamma(\boldsymbol{\lambda})$ function becomes clearly convex in all directions. In practical applications, the numerical minimization
of $\Gamma(\boldsymbol{\lambda})$ would be more efficient.

When enforcing inconsistent data without taking into account experimental error, the behavior is significantly different. Indeed, the only manner to enforce data where the value of the two components of $\boldsymbol{s}$ are different is to significantly displace the two peaks.
On the contrary, the distortion is significantly alleviated when
taking into account experimental errors. Obviously, in this case the
experimental value is not exactly enforced and, with both Gaussian and Laplace prior, we obtain
$\langle \boldsymbol{s} \rangle \approx (0.7,0.7)$.

By observing $\Gamma(\boldsymbol{\lambda})$ it can be seen that the main effect
of using a Laplace prior instead of a Gaussian prior for the error is that
the range of suitable values for $\lambda$ is limited. This allows one to 
decrease the effect of particularly wrong data points on the posterior distribution.

In conclusion, when data that are not consistent among themselves are enforced, the posterior distribution could be severely distorted.
Inconsistency between data could either be explicit (as in the
case where constraints with different reference values are enforced on the same observable) or more subtle. In the reported example, the only way to know that the two components of $\boldsymbol{s}$
should have similar values is to observe their distribution according to the original force field. In the case of complex molecular systems
and of observables that depend non-linearly on the atomic coordinates,
it is very difficult to detect inconsistencies between data points a priori. By properly modeling experimental error it is possible to greatly alleviate the effect of these inconsistencies on the resulting posterior.
Clearly, if the quality of the prior is very poor, correct data points might artificially appear as inconsistent.

\section{Strategies for the Optimization of Lagrangian Multipliers}
\label{sec-strategies}

In order to find the optimal values of the Lagrangian multipliers, one has to minimize the function $\Gamma\left(\boldsymbol{\lambda}\right)$.
The simplest possible strategy is gradient descent (GD), that is an iterative algorithm in which function arguments are adjusted by following the opposite direction of the function gradient.
By using the gradient in Eq.~\ref{gamma-gradient}
the value of $\lambda$ at the iteration $k+1$ can be obtained from the value of $\lambda$ at the iteration $k$ as:

\begin{equation}
\lambda_i^{(k+1)}=\lambda_i^{(k)}-\eta_i\frac{\partial \Gamma}{\partial \lambda_i}=\lambda_i^{(k)}-\eta_i \left(s_i^{exp} -
\langle s_i(\boldsymbol{q})\rangle_{\boldsymbol{\lambda}^{(k)}}
-
\langle \epsilon_i\rangle_{\boldsymbol{\lambda}^{(k)}}
\right)~,
\label{gradient-descent}
\end{equation}
where $\eta$ represents the step size at each iteration and 
might be different for different observables. Here we explicitly indicated that the average
$\langle s_i\left(\boldsymbol{q}\right)\rangle$
should be computed using the
Lagrangian multipliers at the $k^{th}$ iteration $\boldsymbol{\lambda}^{(k)}$.
In order to compute this average it is in principle necessary to sum over
all the possible values of $\boldsymbol{q}$. This is possible for the simple model systems discussed in Section \ref{sec:model-systems},
where integrals can be done analytically. However, for a real molecular system, summing over all the conformations would be virtually impossible.
Below we discuss some possible alternatives.

Notice that this whole review is centered on constraints in the form of Eq.~\ref{eq:maxent}. The methods discussed here can be applied
to inequality restraints as well, as discussed in Ref.~\cite{cesari2016combining}.

\subsection{Ensemble Reweighting}
\label{sec:reweighting}
If a trajectory has been already produced using the prior
force field $V_0(\boldsymbol{q})$, samples from this trajectory might be used
to compute the function $\Gamma(\boldsymbol{\lambda})$. In particular, the 
integral in Eq.~\ref{gamma-func1-noerr} can be replaced by an average over $N_s$ snapshots $\boldsymbol{q}_t$ sampled from $P_0(\boldsymbol{q})$:

\begin{equation}
\tilde\Gamma(\boldsymbol{\lambda})=\ln \left(\frac{1}{N_s}\sum_{t=1}^{N_s} e^{-\boldsymbol{\lambda}\cdot\boldsymbol{s}(\boldsymbol{q}_t)}\right)
+\boldsymbol{\lambda}\cdot \boldsymbol{s}^{exp}
+\Gamma_{err}(\boldsymbol{\lambda})~.
\label{gamma-reweighting}
\end{equation}
A gradient descent on $\tilde{\Gamma}$ results in a procedure
equivalent to
Eq.~\ref{gradient-descent} where the ensemble average
$\langle \boldsymbol{s}(\boldsymbol{q})\rangle_{\boldsymbol{\lambda}^{(k)}}$
is computed as a weighted average on the available frames:

\begin{equation}
\lambda_i^{(k+1)}=\lambda_i^{(k)}-\eta_i\frac{\partial \tilde{\Gamma}}{\partial \lambda_i}=\lambda_i^{(k)}-\eta_i \left(s_i^{exp} -
\frac{\sum_{t=1}^{N_s} s_i(\boldsymbol{q}_t)
e^{-\boldsymbol{\lambda}^{(k)}\cdot\boldsymbol{s}(\boldsymbol{q}_t)}}{\sum_{t=1}^{N_s} e^{-\boldsymbol{\lambda}^{(k)}\cdot \boldsymbol{s}(\boldsymbol{q}_t)}}
-
\langle \epsilon_i\rangle_{\boldsymbol{\lambda}^{(k)}}
\right)~.
\label{gradient-descent2}
\end{equation}
It is also possible to use conjugated gradient or more advanced minimization methods.
Once the multipliers $\boldsymbol{\lambda}^*$ have been found one can compute any
other expectation value by just assigning a normalized weight $ w_t=e^{-\boldsymbol{\lambda}^*\cdot                         \boldsymbol{s}(\boldsymbol{q}_t)} / \sum_{t'=1}^{N_s} e^{-\boldsymbol{\lambda}^*\cdot \boldsymbol{s}(\boldsymbol{q}_{t'})}$
to the snapshot $\boldsymbol{q}_t$.

A reweighting procedure related to this one is at the core of the ensemble-reweighting-of-SAXS method~\cite{rozycki2011saxs}, that has been
used to construct structural ensembles of proteins compatible with
SAXS data~\cite{rozycki2011saxs,boura2011solution}.
Similar reweighting procedures were used to
enforce average data on a variety of systems
\cite{Das2014,sanchez2014application,Hummer2015,leung2015rigorous,BrookesHead-Gordon2016,cunha2017unraveling,bottaro2017conformational,podbevsek2018structural}.
These procedures are very practical since they allow incorporating
experimental constraints a posteriori without the need to repeat the MD simulation.
For instance, in Ref.~\cite{bottaro2017conformational} it was possible
to test different combinations of experimental restraints in order to
evaluate their consistency.
However, reweighting approaches must be used with care since they are effective only
when the posterior and the prior distributions are similar enough \cite{shen2008statistical}.
In case this is not true, the reweighted ensembles will be dominated by a
few snapshots with very high weight, leading to a large statistical error.
The effective number of snapshots with a significant weight can be estimated using
the Kish's effective sample size \cite{GrayKish1969}, defined as
$1/\left(\sum_{t=1}^{N_s}w_t^2\right)$ where $w_t$ are the normalized weights,
or similar measures \cite{martino2017effective}, and is related to the increase of
the statistical error of the averages upon reweighting.

\subsection{Iterative Simulations}

In order to decrease the statistical error, it is convenient
to use
the modified potential
$V(\boldsymbol{q})=V_0(\boldsymbol{q})+k_B T \boldsymbol{\lambda}
\cdot \boldsymbol{s}(\boldsymbol{q})$
to run a new simulation, in an iterative manner. For instance,
in the iterative Boltzmann method, pairwise potentials are modified
and new simulations are performed until the radial distribution function
of the simulated particles does match the desired one
\cite{reith2003deriving}.

It is also possible to make a full optimization of $\Gamma(\boldsymbol{\lambda})$ using a reweighting procedure like the one
illustrated in Section \ref{sec:reweighting} at each iteration.
One would first perform a simulation using the original force field
and, based on samples taken from that simulation, find the optimal
$\boldsymbol{\lambda}$ with a gradient descent procedure. Only at that point a new simulation would be required using a modified potential
that includes the extra $k_B T \boldsymbol{\lambda}
\cdot \boldsymbol{s}(\boldsymbol{q})$ contribution.
This whole procedure should be then repeated until the value of
$\boldsymbol{\lambda}$ stops changing.
This approach was used
in Ref.~\cite{norgaard2008experimental} in order to adjust a
force field to reproduce ensembles of disordered proteins. The same scheme was later used
in a maximum entropy context to enforce average contact maps in the
simulation of chromosomes~\cite{giorgetti2014predictive,tiana2016structural}.
A similar iterative approach was used in Refs.~\cite{zhang2015topology,zhang2016shape}.

In principle, iterative procedures are supposed to converge to the 
correct values of $\boldsymbol{\lambda}$. However, this happens only
if the simulations used at each iteration are statistically converged. %
For systems that exhibit multiple metastable states and are thus difficult to sample it might be difficult to tune the length of each iteration so as to obtain good estimators of the $\Gamma(\boldsymbol{\lambda})$ gradients.

\subsection{On-the-fly Optimization with Stochastic Gradient Descent}\label{sgd-optimization}

Instead of trying to converge the calculation of the gradient at each individual iteration and,
only at that point, modify the potential in order to run a new simulation,
one might try to change the potential
on-the-fly so as to force the system to sample the posterior distribution:

\begin{equation}
V(\boldsymbol{q},t)=V_0(\boldsymbol{q})+k_B T \boldsymbol{\lambda}(t)
\cdot \boldsymbol{s}(\boldsymbol{q})~.
\end{equation}
An earlier approach aimed at enforcing time-averaged constraints was reported in Ref.~\cite{TordaScheekGunsteren1989}.
However, here we will focus on methods based on the maximum-entropy formalism.

The simplest choice in order to minimize the $\Gamma(\boldsymbol{\lambda})$ function is to
use a 
stochastic
gradient descent (SGD) procedure, where an unbiased estimator of the gradient is used to update $\boldsymbol{\lambda}$.
In particular, the instantaneous value of the forward model computed at time $t$, that is $\boldsymbol{s}(\boldsymbol{q}(t))$,
can be used to this aim.
The update rule for $\boldsymbol{\lambda}$ can thus be rewritten as a differential equation:

\begin{equation}
\label{eq:update-continuous}
\dot{\lambda}_i(t)=-\eta_i \left(t\right) \left(s_i^{exp} -
s_i(\boldsymbol{q}(t))
-\langle \epsilon_i \rangle_{\boldsymbol{\lambda}_i(t)}
\right)
\end{equation}
with initial condition $\boldsymbol{\lambda}(0)=\boldsymbol{0}$.

Notice that now $\eta$ plays the role of a \emph{learning rate} and depends on the simulation time $t$. This choice is motivated by the fact that approximating the true gradient with its unbiased estimator introduces a noise into its estimate. In order to decrease the effect of such noise, a common choice when using SGD is to reduce the learning rate as the minimization (learning) process progresses with a typical schedule $\eta(t)\propto 1/t$ for large times.
In our previous work \cite{cesari2016combining} we adopted a learning rate from the class \textit{search then converge} \cite{darken1991}, which prescribes to choose $\eta_i \left(t\right)=k_i/\left(1+\frac{t}{\tau_i}\right)$. Here $k_i$ represents the initial learning rate and $\tau_i$ represents its damping time. In this manner, the learning rate is large at the beginning of the simulation and decreases proportionally to $1/t$ for large simulation times. The parameters $k_i$ and $\tau_i$ are application specific and must be tuned by a trial and error procedure. In particular, a very small value of $\tau$ will cause the learning rate to decrease very fast, increasing the probability to get stuck in a suboptimal minimum. On the other hand, a very large value of $\tau$ will prevent step-size shrinking and thus will hinder convergence. Analogous reasoning also applies to $k$ (see Section~\ref{sec:langevin} for numerical examples).
Also notice that the $k_i$'s are measured in units of the inverse of the observable squared multiplied by an inverse time
and could thus in principle be assigned to different values in case of heterogeneous observables.
It appears reasonable to choose them inversely proportional to the observable variance in the prior, in order to make
the result invariant with respect to a linear transformation of the observables.
On the other hand, the $\tau_i$ parameter should probably be independent of $i$ in order to avoid
different $\lambda_i$'s to converge on different timescales.

Once Lagrangian multipliers are converged or, at least, stably fluctuating around a given value, the optimal value $\boldsymbol{\lambda}^*$ can be estimated by taking a time
average of $\boldsymbol{\lambda}$ over a suitable time window.
At that point, a new simulation could be performed using
a static potential
$
V^*(\boldsymbol{q})=V_0(\boldsymbol{q})+k_B T \boldsymbol{\lambda}^*
\cdot \boldsymbol{s}(\boldsymbol{q})
$, either from a different molecular structure or starting from the
structure obtained at the end of the learning phase.
Such a simulation done with a static potential can be used to
rigorously validate the obtained $\boldsymbol{\lambda}^*$.
Notice that, if errors have been included in the model, such validation
should be made by checking that $\langle \boldsymbol{s}\rangle
\approx \boldsymbol{s}^{exp}-\langle \boldsymbol{\epsilon} \rangle
$. Even if the resulting $\boldsymbol{\lambda}^*$ are suboptimal, it is plausible that such a simulation could be further reweighted (Section~\ref{sec:reweighting}) more easily than the one performed with the original force field.
When modeling errors,
if an already restrained trajectory is reweighted one should be aware that restraints will be overcounted
resulting in an effectively decreased experimental error (see Section~\ref{sec:modeling-errors}).

As an alternative, one can directly analyze the learning simulation.
Whereas strictly speaking this simulation is performed out of equilibrium, this approach has the advantage that
it allows the learning phase to be prolonged until the agreement with experiment is satisfactory.

The optimization procedure discussed in this Section was used in order to enforce NMR data on RNA nucleosides and dinucleotides  in Ref.~\cite{cesari2016combining}, where it was further extended in order to
simultaneously constrain multiple systems by keeping their
force fields chemically consistent. This framework represents a promising avenue for the improvement of force fields, although it is
intrinsically limited by the fact that the functional form of the correcting potential is by construction related to the type
of available experimental data. However, the method in its basic
formulation described here can be readily used in order to enforce system-specific experimental constraints.

Finally, notice that Eq.~\ref{eq:update-continuous} is closely related to the on-the-fly procedure proposed in the appendix of Ref.~\cite{Hummer2015},
where a term called ``generalized force'' and proportional to $\lambda$ is calculated from an integral over the trajectory.
Using the notation of this review, considering a Gaussian prior for the error (Section \ref{sec:modeling-errors}),
and setting the confidence in the force field $\theta=1$, 
the time-evolution of $\lambda$ proposed in Ref.~\cite{Hummer2015} could
be rewritten in differential form as Eq.~\ref{eq:update-continuous} with $\eta(t)=1/(\sigma_i^2t)$,
however with a different initial condition $\boldsymbol{\lambda}(0)=\boldsymbol{s}(\boldsymbol{q}(0))/\sigma_i^2$.

\subsection{Other On-the-fly Optimization Strategies}
\label{sec:other-on-the-fly}

Other optimization strategies have been proposed in the literature.
The already mentioned target metadynamics (Section \ref{sec:enforcing-distributions}) provides a framework to enforce experimental data,
and was applied to enforce reference distributions obtained from
more accurate simulation methods \cite{white2015designing},
from DEER experiments
\cite{marinelli2015ensemble}, or from conformations collected over
structural databases \cite{Gil-LeyBottaroBussi2016}. 
It is however not clear if it can be extended to enforce individual averages.

Also the VES method \cite{Valsson2014} (Section \ref{sec:enforcing-distributions})
is designed to enforce full distributions. However, in its practical implementation, the correcting potential is expanded on a basis set and the average values of the basis functions are actually constrained, resulting thus numerically equivalent to the other methods discussed here.
In VES, a function equivalent to $\Gamma(\boldsymbol{\lambda})$
is optimized using the algorithm by Bach and Moulines \cite{bach2013non} that is optimally suitable for non-strongly-convex functions.
This algorithm requires to estimate not only the gradient but also the Hessian of the function $\Gamma(\boldsymbol{\lambda})$. We recall that $\Gamma(\boldsymbol{\lambda})$
can be made strongly convex by suitable treatment of experimental
errors (see Section \ref{sec:modeling-errors}). However, there might be situations where the Bach-Moulines algorithm outperforms the SGD.

The experiment-directed simulation (EDS) approach \cite{WhiteVoth2014} instead
does not take advantage of the function $\Gamma(\boldsymbol{\lambda})$ but rather minimizes
with a gradient-based method \cite{duchi2011adaptive} the square deviation between the experimental values and
the time-average of the simulated ones.
A later paper tested a number of related minimization strategies~\cite{hocky2017coarse}.
In order to compute the gradient of the ensemble averages
$\langle s_i \rangle_{\boldsymbol{\lambda}}$ with respect to $\boldsymbol{\lambda}$ it is necessary to compute the variance of
the observables $s_i$ in addition to their average.
Average and variance are computed on short simulation segments.
It is worth observing that obtaining an unbiased estimator 
for the variance is not trivial if the simulation segment is too short.
Errors in the estimate of the variance would anyway only affect the effective learning rate of the Lagrangian multipliers.
In the applications performed so far, a few tens of MD time steps were shown to be sufficient to this aim,
but the estimates might be system dependent.
A comparison of the approaches used in Refs.~\cite{WhiteVoth2014,hocky2017coarse} with the SGD proposed in Ref.~\cite{cesari2016combining} in practical applications would be useful to better understand the pros and the cons
of the two algorithms.
EDS was used to enforce the gyration radius of a 16-bead polymer to match the one of a reference system \cite{WhiteVoth2014}.  Interestingly, the restrained polymer was reported to have not only the average gyration radius in agreement with the reference one, but also its distribution.
This is a clear case where a maximum entropy (linear) restraint and a harmonic restraint give completely different results.
The EDS algorithm was recently applied to a variety of systems
(see, e.g., Refs.~\cite{DannenhofferLafage2016,hocky2017coarse,white2017communication}).

\section{Convergence of Lagrangian Multipliers in Systems Displaying Metastability}

Evaluating the Lagrangian multipliers on-the-fly might be nontrivial especially in systems that present multiple
metastable states. We here present some example using a model system and provide some recommendation for the usage of enhanced sampling methods.

\subsection{Results for a Langevin System}
\label{sec:langevin}
We first illustrate the effect of the choices in the
learning schedule on the convergence of the Lagrangian multipliers
and on the sampled distribution
when using a SGD approach (Section \ref{sgd-optimization}). We consider a one dimensional system
subject to a potential
$V_0(s)=-k_BT\ln\left(
e^{-(s-s_A)^2/2\sigma^2}+e^{-(s-s_B)^2/2\sigma^2}
\right)$, with $s_A=0$, $s_B=3$, and $\sigma=0.4$.
The system is evolved according to an overdamped
Langevin equation with diffusion coefficient $D=1$ using a timestep $\Delta t=0.01$.
The average value of $s$ in the prior distribution is $\langle s \rangle_0=(s_A+s_B)/2=1.5$.
The potential has been chosen in order to exhibit a  free-energy barrier and is
thus representative of complex systems where multiple metastable states are available.

\begin{table}
\caption{
\label{table}
Summary of the results obtained with the Langevin model, including
learning parameters ($k$ and $\tau$) and average $\langle\lambda\rangle$
and $\langle s\rangle$ computed over the second half of the simulation.
In addition, we report the exact Lagrangian multiplier $\lambda^*_{\langle s\rangle}$ required to enforce
an average equal to $\langle s\rangle$ and the exact average $\langle s\rangle_{\langle\lambda\rangle}$
corresponding to a Lagrangian multiplier $\langle\lambda\rangle$. The last two columns are obtained
by using the analytical solutions described in Section \ref{sec:model-systems}.
Panel labels match those in Fig.~\ref{fig-langevin}.
}
\begin{center}
\begin{tabular}{c|cccccc}
$\texttt{Panel}$  & $k$  & $\tau$  & $\langle\lambda\rangle$  & $\langle s\rangle$  & $\lambda^*_{\langle s\rangle}$  & $\langle s\rangle_{\langle\lambda\rangle}$\tabularnewline
\hline 
$\texttt{a}$  & $2$  & $10$  & $0.207$  & $1.008$  & $0.210$  & $1.015$\tabularnewline
$\texttt{b}$  & $2$  & $0.001$  & $0.080$  & 1.308  & $0.080$  & $1.307$\tabularnewline
$\texttt{c}$  & $0.001$  & $10$  & $0.077$  & $1.324$  & $0.073$  & $1.316$\tabularnewline
$\texttt{d}$  & $2$  & $10000$  & $0.145$  & $1.000$  & $0.214$  & $1.157$\tabularnewline
$\texttt{e}$  & $1000$  & $10$  & $0.158$  & $1.001$  & $0.214$  & $1.125$\tabularnewline
\end{tabular}
\end{center}
\end{table}

\begin{figure}
\centering
\includegraphics[width=\linewidth]{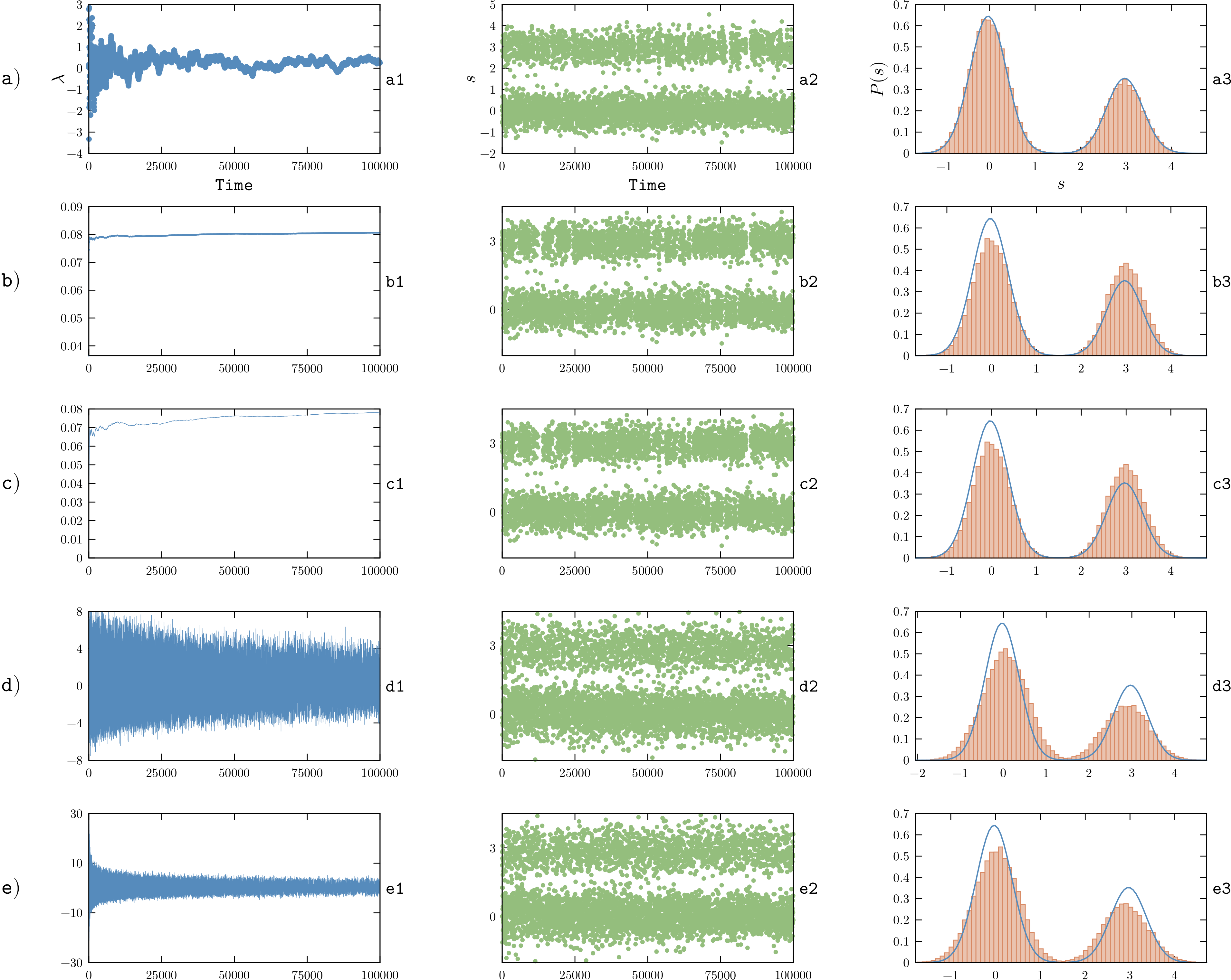}
\caption{Effect of choosing different values for $k$ and $\tau$ when using SGD on-the-fly during MD simulations. In particular, for each set of $k$ and $\tau$ we show the convergence of the Lagrangian multipliers (number $1$ of each letter), the time series of the observable (number $2$ of each letter), and the resulting sampled posterior distribution, red bars,
together with the analytical result, continuous line (number $3$ of each letter). Panel labels match those of Table~\ref{table}.
}
\label{fig-langevin}
\end{figure}

We then
run an on-the-fly SGD scheme \cite{cesari2016combining} in order to enforce an experimental average $s^{exp}=1$.
For simplicity, experimental error is not modeled. By using the
analytical results of Section~\ref{sec:model-systems}, it can be seen that the exact Lagrangian multipliers required to enforce this average
is $\lambda^*=0.214$.
In particular, we test different choices for $k$ and $\tau$ which represent respectively the initial value of the learning rate and its damping factor (see \ref{sgd-optimization} for more details on these parameters).
The list of parameters and the results are summarized in Table~\ref{table}, whereas
Fig.~\ref{fig-langevin} reports the actual trajectories, their histogram, and the time evolution of the Lagrangian multipliers.

Panels \texttt{a1} and \texttt{a3} in Fig.~\ref{fig-langevin} report results obtained with a correct choice of the parameters. Panel \texttt{a1} shows that the Lagrangian multiplier has quite large fluctuations at the beginning of the simulation (as expected from SGD), which are then damped as the simulations proceeds. The resulting sampled posterior distribution (red bars in panel \texttt{a3}) is in close agreement with the analytical solution (continuous blue line). The resulting average $\langle \lambda \rangle\approx 0.207$ reported in Table~\ref{table} is
in very good agreement with the analytical result ($\lambda^*=0.214$).

Panels \texttt{b1} and \texttt{b3} in Fig.~\ref{fig-langevin} show the effect of choosing a very small value of $\tau$. This choice not only kills the noise but also hinders the convergence of $\lambda$ by shrinking too much the step-size during the minimization. The resulting distribution shown in panel \texttt{b3} is clearly in disagreement with the analytical one having wrong populations for the two peaks. This example shows that apparently converged Lagrangian multipliers (panel \texttt{b1}) are not a sufficient condition for convergence to the correct result, and it is necessary to check that the correct value was actually enforced. Panels \texttt{c1} and \texttt{c3} in Fig.~\ref{fig-langevin} show the effect of choosing a too small value of $k$. This scenario is very similar to the previous one since both cases result in small values of the learning rate $\eta$. Thus, what said for \texttt{b1} and \texttt{b3} also applies to \texttt{c1} and \texttt{c3}. As reported in Table~\ref{table}, in both cases the final average is $\langle s\rangle\approx 1.3$ and is thus visibly different from $s^{exp}=1$. Thus, in a real application, this type of pathological behavior would be easy to detect. We recall that in case error is explicitly modeled
(Section \ref{sec:modeling-errors}) one should compare
$\langle s \rangle$ with $s^{exp}-\langle \epsilon \rangle_{\lambda}$.

Panels \texttt{d1} and \texttt{d3} in Fig.~\ref{fig-langevin} show the effect of choosing a very large value of $\tau$. The effect of such choice is that the damping rate of the noise in Lagrangian multipliers is much slower than in the ideal case. This is reflected in the larger fluctuations of Lagrangian multipliers (panel \texttt{d1}) but also in an incorrect reconstruction of the posterior.
The last example, panels \texttt{e1} and \texttt{e3} in Fig.~\ref{fig-langevin}, shows the effect of choosing a very large value of $k$. In this case, the fluctuations of the Lagrangian multiplier (panel \texttt{e1}) are even higher than in the previous case. As reported in Table~\ref{table}, in both cases the final average is equal to $s^{exp}=1$.
So, even though
the sampled distribution has the correct average
it is not the distribution that maximizes the entropy.
This is a suboptimal solution that might be
at least qualitatively satisfactory in some case.
However, it is clear that there is no way to detect
the incorrectness in the resulting distribution
by just monitoring the enforced average. The only practical way to detect the problem indeed is to consider the resulting value of $\langle \lambda \rangle\approx 0.15$ and run a new simulation with a static potential. An additional indication of the problematic
behavior is the large (several units) fluctuations in the
Lagrangian multipliers. Indeed, the problem can be rationalized noting that the
timescale at which $\lambda$ evolves is too fast when compared with the typical time required to see a transition between one state and the other and the restraining force is overpushing the system forcing it to spend too much time in the region between the two peaks.
The problem can be solved either slowing down the $\lambda$ evolution (as in panel \texttt{a}) or by using enhanced sampling methods to increase the number of transitions.

\subsection{Comments about Using Enhanced Sampling Methods}

The model potential discussed above displays a free-energy barrier separating two metastable states.
In order to properly sample both peaks in the distribution it is necessary to wait the  time required
to cross the barrier. If the transition is forced by very large fluctuations of $\lambda$,
one can see that the resulting distribution is significantly distorted. For this reason, whenever a system
displays metastability, it is highly recommended to use enhanced sampling techniques~\cite{bernardi2015enhanced,valsson2016enhancing,mlynsky2017exploring}.
It is particularly important to employ methods that are capable to induce transitions between the states that contribute
to the measured experimental averages.
NMR timescales of typically $\mu s-ms$, i.e., upper limits for the lifetimes of interconverting conformations that are 
indistinguishable in the spectra, can be reached better using enhanced sampling techniques,  
since they result in probability distributions that would be effectively sampled by a much longer un-enhanced continuous simulation. 

Replica exchange methods where one replica is unbiased are easy to apply since 
the learning procedure can be based on the reference replica.
Methods such as parallel tempering~\cite{sugita1999replica}, solute tempering~\cite{liu2005replica},
bias-exchange metadynamics with a neutral replica~\cite{piana2007bias},
or collective-variable tempering~\cite{Gil-LeyBussi2015}
can thus be used straightforwardly. Notice that in this case the higher replicas might feel either the same
correcting potential as the reference replica
(as it was done in Ref.~\cite{cesari2016combining}) or might be independently subject to the experimental restraints, provided
the differences in the potential energy functions are properly taken into account in the acceptance rate calculation.
Leaving the higher replicas uncorrected (i.e., simulated with the original force field) is suboptimal since they would explore a different portion of the space leading to fewer exchanges with the reference replica. 
It is also important to consider that, thanks to the coordinate exchanges, the reference replica will be visited by different conformations. These multiple conformations will all effectively contribute to the
update of the Lagrangian multipliers. For instance,
if a SGD is used, in the limit of very frequent exchanges, the update will be done according to the average value of the observables over the conformations of all replicas, properly weighted with their probability to visit the reference replica.

Methods based on biased sampling,
such as umbrella sampling \cite{torrie1977nonphysical},
metadynamics
\cite{laio2002escaping},
parallel-tempering metadynamics \cite{bussi2006free},
bias-exchange (without a neutral replica) \cite{piana2007bias}
or parallel-bias \cite{pfaendtner2015efficient} metadynamics, require instead the implementation of some on-the-fly reweighting procedure in order to properly perform the update
of the Lagrangian multipliers. The
weighting factors should be proportional to the exponential of the biasing potential and, as such, might lead to a very large variability
of the increment of $\lambda$ (Eq.~\ref{eq:update-continuous}) that
could make the choice of the learning parameters more difficult.
For instance, this could result in very large $\lambda_i$'s in the initial transient leading to large forces that make the simulation
unstable.
When reweighting (see Section~\ref{sec:reweighting})
a trajectory generated using one of these enhanced sampling methods
it is sufficient to use the weighting factors
in the evaluation of Eq.~\ref{gamma-reweighting}.
Notice that similar arguments apply to replica-based methods (see Section~\ref{sec:replica}), where on-the-fly reweighting is required in order to correctly
compute the replica average~\cite{bonomi2016metadynamic}. 
If the resulting weights of different replicas are too different, the average value might be dominated by a single or a few replicas.
A low number of replicas contributing to the average might in turn lead to large forces, unless the spring constant is suitable reduced
\cite{lohr2017metadynamic}, and to an entropy decrease.

Even in the absence of any enhanced sampling procedure, we notice that Lagrangian multipliers could be updated
in a parallel fashion by multiple equivalent replicas, in a way resembling that used in multiple-walkers
metadynamics to update the bias potential~\cite{raiteri2006efficient}.
Since $s_i(\boldsymbol{q})$ enters linearly in Eq.~\ref{eq:update-continuous}, this would be totally equivalent
to using the arithmetic average between the walkers to update the Lagrangian multipliers
(to be compared with the weighted average discussed above for replica-exchange simulations), showing an interesting
analogy between Lagrangian multiplier optimization and replica-based methods (see Section~\ref{sec:replica}).
Such a multiple-walkers approach was used for instance in the well-tempered variant of
VES~\cite{valsson2015well}, although in the context of enhanced sampling rather than to enforce experimental data.

\section{Discussions and Conclusions}

In this work we reviewed a number of recently introduced techniques
that are based on the maximum entropy principle and that allow experimental observations to be incorporated in MD simulations preserving the
heterogeneity of the ensemble. We here discuss some general features of the reviewed methods.

First, one must keep in mind that, by design, the maximum entropy principle provides a distribution that, among those satisfying the experimental constraints, is as close as possible to the prior distribution. 
If the prior distribution is reasonable, a minimal correction is expected to be a good choice. However, for systems where the performance of classical force fields is very poor, the maximum entropy principle should be used with care and, if possible, should be based on a large number of experimental data so as to diminish the impact of force-field deficiencies on the final result.
As a related issue, different priors are in principle expected to lead to different posteriors, and thus different ensemble averages
for non-restrained quantities. There are indications that current
force fields restrained by a sufficient number of experimental data points lead to equivalent posterior distributions
at least for trialanine \cite{Das2014} and for larger disordered peptides~\cite{tiberti2015encore,lohr2017metadynamic}.
It would be valuable to perform similar tests on other systems where
force fields are known to be poorly predictive, such as unstructured RNAs or difficult-to-predict RNA structural motifs.

We here discussed both the possibility of reweighting a posteriori
a trajectory and that of performing a simulation where the restraint
is iteratively modified.
Techniques where the elements of a previously generated ensemble are reweighted  have the disadvantage that if the initial ensemble averages are far from the experimental values the weights will be distributed very inhomogeneously (i.e., very large $\lambda_i$ will be needed), which means that singular conformations with observables close to the experimental values can be heavily overweighted to obtain the correct ensemble average.
In the extreme case, it might not even be possible to find weights that satisfy the desired ensemble average, since important conformations are simply missing in the ensemble.
On the other hand, reweighting techniques have the advantage that they can be readily applied to new or different experimental data, without performing new simulations.
Additionally, they can be used to reweight a non-converged simulation performed with an on-the-fly optimization.

When simulating systems that exhibit multiple metastable states,
it might be crucial to combine the experimental constraints with
enhanced sampling methods. This is particularly important if 
multiple metastable states contribute to the experimental average.
As usual in enhanced sampling simulations, one should observe as many as possible transitions between the relevant metastable states.
When using replica-based methods (either to enhance sampling or to compute averages), transitions should be observed in the continuous trajectories.

Several methods are based on the idea of simulating a number of replicas of the system with a restraint on the instantaneous average among the replicas and have been extended to treat experimental errors. These methods are expected to reproduce the maximum entropy distribution in the limit of a large number of replicas. However, if the number of replicas is too low, the deviation from the maximum entropy distribution might be significant. Indeed, the number of replicas should be large enough for all the relevant states to be represented with the correct populations. The easiest way to check if the number of replicas is sufficient is to compare simulations done using a different number of replicas. Methods based on Lagrangian multipliers reproduce the experimental averages by means of an average over time rather than an average over replicas. Thus, they can be affected by a similar problem if the simulation is not long enough.
This sort of effect is expected to decrease when the simulation length increases and when using enhanced sampling techniques.

The on-the-fly refinement of Lagrangian multipliers typically requires ad hoc parameters for the learning phase that should be chosen
in a system-dependent manner. Properly choosing these parameters is not trivial.
Several different algorithms have been proposed in the last years and a systematic comparison
on realistic applications would be very useful. It might also be beneficial to consider other stochastic optimization
algorithms that have been proposed in the machine-learning community.
Interestingly, all the methods discussed in this review for on-the-fly optimization (target metadynamics, maximum entropy with SGD, VES, and EDS) are available in the latest release of the software
PLUMED~\cite{PLUMED2_2014} (version 2.4), which also implements replica-based methods,
forward models to calculate experimental
observables~\cite{bonomi2017integrative},
and enhanced sampling methods.

Finally, we notice that there are cases where results might be easier to interpret if only a small number of different conformations were contributing to the experimental average.
In order to obtain small sets of conformations that represent the ensemble and provide a clearer picture about the different states, several \textit{maximum parsimony} approaches have been developed.
Naturally, the selection of a suitable set of structures is done on an existing ensemble and not on-the-fly during a simulation.
While some approaches use genetic algorithms to select the structures of a fixed-size set \cite{Bernado2007,Nodet2009,Pelikan2009}, others use matching pursuit \cite{Berlin2013} or Bayes-based reweighting techniques to obtain correct ensemble averages \cite{Yang2010,Fisher2010,Cossio2013,Molnar2014}  while minimizing the number of non-zero weights, i.e. structures, in the set.
These approaches are not central to this review and so were not
discussed in detail.

In conclusion, the maximum (relative) entropy principle provides a consistent framework to combine molecular dynamics simulations and experimental data. On one hand, it allows improving not-satisfactory results sometime obtained when simulating complex systems with classical force fields. On the other hand, it allows the maximum amount of structural information to be extracted from experimental data, especially in cases where heterogeneous structures contribute to a given experimental signal.
Moreover, if experimental errors are properly modeled, this framework
allows to detect experimental data that are either mutually inconsistent or incompatible with the employed force field.
For all these reasons, we expect this class of methods to be increasingly applied for the characterization of the structural dynamics of biomolecular systems in the coming future.

\vspace{6pt} 

\acknowledgments{
Max Bonomi, Sandro Bottaro, Carlo Camilloni, Glen Hocky, Gerhard Hummer, Juergen Koefinger, Omar Valsson, and Andrew White are acknowledged for reading the manuscript and providing useful suggestions. Kresten Lindorff-Larsen is also acknowledged for useful discussions.
}

\conflictofinterests{The authors declare no conflict of interest.}
\abbreviations{The following abbreviations are used in this manuscript:\\

\noindent
DEER: double electron-electron resonance\\
EDS: experiment-directed simulation\\
GD: gradient descent\\
MD: molecular dynamics\\
NMR: nuclear magnetic resonance\\
SAXS: small-angle X-ray scattering\\
SGD: stochastic gradient descent\\
VES: variationally enhanced sampling
}

\bibliographystyle{mdpi}

\bibliography{bibliography}

\end{document}